\documentclass[3p,onecolumn]{elsarticle}
\usepackage{todonotes}
\usepackage{longtable}
\usepackage[latin1]{inputenc}
\usepackage[T1]{fontenc}
\usepackage[english]{babel}
\usepackage{amsmath}
\usepackage{amssymb,amsfonts,textcomp}
\usepackage{array}
\usepackage{supertabular}
\usepackage{hhline}
\usepackage{graphicx}
\usepackage{tabularx}	
\usepackage{enumitem}
\usepackage{soul}
\usepackage{url}

\makeatletter
\newcommand\arraybslash{\let\\\@arraycr}
\newcommand {\challengesnum}{11~}
\newcommand{\rqone}{what challenges do practitioners face in integrating UX practices into software development processes?}
\newcommand{\rqtwo}{how do these challenges relate to challenges in practice of software quality characteristics, in particular usability?}
\newcommand {\intervieweesnum}{17~}
\newcommand {\issuesnum}{five~}
\newcommand {\difficultiesnum}{seven~}
\newcounter{rchlno}
\DeclareRobustCommand{\rchl}[1]{%
   \refstepcounter{rchlno}%
   \therchlno\label{#1}}

\newcommand{\libpath}{/Users/puix/pariya/research/library.bib}

\makeatletter
\def\ps@pprintTitle{%
	\let\@oddhead\@empty
	\let\@evenhead\@empty
	\def\@oddfoot{}%
	\let\@evenfoot\@oddfoot}
\makeatother
\makeatother

\title{Integrating User eXperience Practices into Software Development Processes:\\ Implications of Subjectivity and Emergent Nature of UX}

\begin{document}

	\begin{frontmatter}
		\begin{abstract}
				User eXperience (UX) is a key factor in the success of software systems.
				Many software companies face challenges in their work with UX and how to integrate UX practices into existing development processes.
				A better understanding of these challenges, based on empirical data, can help researchers and practitioners better address them in the future.
				Existing research does not analyse UX practices and challenges in relation to other software quality characteristics or, in particular, in relation to usability.
				In this empirical study, we have interviewed \intervieweesnum practitioners with different backgrounds and occupations from eight software development companies.
				Their responses are coded, and analysed with thematic analysis. 
				We report \challengesnum challenges that practitioners face in their work with UX.
				Some of these challenges partly overlap with those reported in existing literature about usability or software quality characteristics.
				In contrast to these overlaps, the participants of our study either view many of the challenges unique to UX, or more severe than for usability or other quality characteristics.
				Although at a superficial level challenges with UX and other quality characteristics overlap, we differentiate these challenges at a deeper level through two main aspects of UX: \textit{subjectivity} and \textit{emergent nature}.
				In particular, we identify at least \issuesnum issues that are essential to the very nature of UX, and add at least \difficultiesnum extra difficulties to the work of practitioners.
				These difficulties can explain why practitioners perceive the challenges to be more severe than for other quality characteristics.
				Our findings can be useful for researchers in identifying new and industrially relevant research areas and  for practitioners to learn from empirically investigated challenges in UX work, and base their improvement efforts on such knowledge.
				Investigating the overlaps can help finding research areas not only for enhancing practice of UX but also software quality in general.
				It also makes it easier for practitioners to spot, better understand as well as find mitigation strategies for UX, through learning from past experiences and developments in the area of software quality.
				
		\end{abstract}

		\title{Integrating User eXperience Practices into Software Development Processes:\\ The Implication of Subjectivity and Emergent Nature of UX}
		
	\author{Pariya Kashfi\fnref{fn1}}
		\ead{p@puix.org}
		\fntext[fn1]{(Corresponding author) Ph.Lic, PhD candidate, Department of Computer Science and Engineering,	Chalmers University of Technology, Gothenburg, Sweden}
				
		\author{Agneta Nilsson\fnref{fn2}}
		\ead{agneta.nilsson@chalmers.se}
		\fntext[fn2]{Department of Computer Science and Engineering,	Chalmers University of Technology and Gothenburg University, Gothenburg, Sweden}

\author{Robert Feldt\fnref{fn3}}
\ead{robert.feldt@chalmers.se}
\fntext[fn3]{Department of Computer Science and Engineering,	Chalmers University of Technology, Gothenburg, Sweden}

\end{frontmatter}
			
%
%


\section{Introduction}
As the software industry has matured, the demands that society puts on the quality of software systems has increased. 
It is no longer enough to focus only on the many functions that a piece of software should supply. 
To deliver a system that is consistent and of high quality there are a large number of characteristics that need to be considered~\cite{Chung2000a}. 
Some, such as testability, are internal or relate to the development process and mainly concern developers, while others such as performance and usability, are critical for users~\cite{ISO250102011}. 
At an even broader level, the actual experience of the end users as they interact with the software needs to be taken into account.

Recently this widening scope of software quality characteristics has led to the introduction and study of the concept of user experience (UX). 
Even though different definitions of UX exist they share the same essence: 
\textit{UX is a user's holistic perception of functionalities and quality characteristics of a piece of software}~\cite{Hassenzahl2003a,Wright2010a, Jordan2002a}.
In general, UX literature emphasizes that assuring efficiency and effectiveness during use of the software, i.e high usability, does not guarantee that the end users will have a positive experience~\cite{Hassenzahl2010a}.
A good UX typically means that the software has high usability, but the latter does not automatically lead to the former.

All software systems deliver some UX, positive or not, whether the UX has explicitly been taken into account during development or not.
Research has shown that certain practices can increase the likelihood of delivering a positive UX~\cite{Hassenzahl2010a} (hereafter, \textit{UX practices}).
But simply applying these practices in isolation is not enough~\cite{Abrahao2010a,Ferreira2012a,Ovad2015a}.
Like methods and practices used to support other software quality characteristics~\cite{Chung2000a}, they need to be integrated into development processes and considered throughout projects. 

Despite the recognized importance of integrating UX practices into software development processes, practitioners still lack enough support to enable this integration~\cite{Abrahao2010a,Isomursu2012a}.
As a result, UX practices are often neglected in software projects~\cite{Law2014a}.
The same goes for software quality characteristics~\cite{Chung2009a,Paech2004a,BerntssonSvensson2012a}.
Therefore, there have been calls for more empirical research on the practice of UX~\cite{Abrahao2010a} in line with similar calls concerning other software quality characteristics~\cite{Paech2004a}.

One way to improve state-of-practice is to gain a better understanding of current challenges that software engineers face in their everyday work.
Some researchers therefore have reported challenges concerning the practice of software quality characteristics in general~\cite{Chung2009a,BerntssonSvensson2012a,Borg2003a}, and usability in particular~\cite{Rosenbaum2000a,Boivie2003b,Jerome2005a,Heiskari2009a,Gulliksen2004a}.
There have also been a few studies that directly or indirectly report on challenges practitioners face when applying UX practices (hereafter, \textit{UX challenges}).

We identified three types of these studies.
First, studies that focus on only some aspect of UX work, i.e., measurability of UX (e.g.,~\cite{Law2014b,Gerea2015a}), UX evaluation (e.g.,~\cite{Alves2014a,Vermeeren2010a}), or practitioners' understanding of the concept of UX (e.g.~\cite{Lallemand2014a})
These studies, however essential in providing a better understanding of UX state of practice, do not cover all aspects of UX~\cite{Hassenzahl2010} and thus cannot fully consider the interplay between multiple aspects.
Secondly, there are studies that investigate UX practices in the context of agile projects (e.g.~\cite{Isomursu2012a,Larusdottir2012a,Law2015a}).
These studies often concern various aspects of UX work, but specifically investigate how the setup in one particular type of software project (agile) facilitates or inhibits the practices.
Thirdly, we found studies that claim to investigate UX state of practice, but tend not to differentiate usability and UX and how similarities and differences between these two concepts affect the practice (either in agile development processes~\cite{Ovad2015a,Cajander2013a}, or in software development in  general~\cite{Ardito2014a,Ardito2014b,Lanzilotti2015a}).
The findings of such studies therefore do not necessarily provide sufficient understanding of UX challenges, since UX practices go beyond usability practices~\cite{Abrahao2010a,Law2014a}.
In addition, to the best of our knowledge, none of the current studies on UX challenges have analyzed the implications of similarities and differences between UX and other software quality characteristics.
Since both practitioners and researchers have more experience working with other quality characteristics, practice can be better improved if UX challenges are related to existing knowledge rather than primarily portrayed as wholly new and/or different.

Thus, although there have been calls for more empirical research on UX practices, and some studies have investigated UX challenges, a more complete picture and understanding of UX challenges in different types of software development processes and in relation to other software quality characteristics is missing.
This motivates our study in which eight software organizations with different levels of maturity regarding UX and different development processes participated.
Here we report our findings and answer the following research questions:
\textit{\rqone}, 
and \textit{\rqtwo}
Although we focused on UX challenges, the insights that can be gained from our findings may also shed light on the practice of software quality characteristics in general.
Most importantly, we have discussed these challenges in relation to the \textit{subjectivity} and \textit{emergent nature} of UX, and have analyzed the related existing approaches and open problems for future research. 
The subjectivity and emergent nature of UX resembles two known aspects of software quality characteristics respectively: \textit{subjectivity} and their often \textit{cross-cutting} nature.
We therefore specifically discuss the differences and similarities between these aspects in the case of UX compared to  other quality characteristics.

The structure of this paper is as follows:
The second section explores background and summarizes the related literature.
The third section describes our research methodology and presents the different research sites.
The fourth section presents the results from our study, the identified challenges.
The fifth section discusses our findings, connects these with the related literature, and the research questions.
In the last section, we conclude our study and suggest future research.

\section{Background and Related Work}
\label{sec:bg}

There are a large number of software quality characteristics that practitioners are recommended to take into account in development~\cite{Chung2000a}. 
One such quality characteristic of critical importance for the end users is \textit{usability}~\cite{ISO250102011}. 
ISO/IEC~9241-11~\cite{ISO9241} defines usability as \textit{``the extent to which a product can be used by specified users to achieve specified goals with effectiveness, efficiency and satisfaction in a specified context of use.''}
In this definition, `user satisfaction' refers to the effectiveness and the efficiency of the user's interaction with the software, and focuses on how the user perceives the outcome of this interaction concerning achievement of a goal.
However, more recent research highlights that the users' overall judgment of software is not merely influenced by how they perceive achievement of their goals.
The judgment is also influenced by how users perceive satisfaction of their personal needs such as `being stimulated', `gaining pleasure', or `feeling connected to their loved ones'~\cite{Wright2010a}.
Therefore, to improve our understanding of users' perception of products and services, researchers have introduced the concept of User eXperience (UX)~\cite{Hassenzahl2003a,Jordan2002a}.

Even though different definitions of the concept of UX exist, they share the same essence: 
\textit{UX is users' holistic perception of functionalities and quality characteristics of a piece of software}~\cite{Hassenzahl2003a, Jordan2002a}.
The different definitions reflect a difference in how researchers approach the modeling of UX.
These  approaches mainly differ in their view on how various underlying elements and processes contribute in forming the end user's experience~\cite{Hassenzahl2006b}.
At least four main approaches to the modeling of UX can be found:

\begin{itemize}
	\item  \textit{The experiential}:
	Models in this approach (e.g.~\cite{Wright2010a,Forlizzi2004}) emphasize the `holistic nature' of UX.
	They highlight that any experience is a unique combination of complexly interrelated, inseparable elements, e.g., users' expectations and properties of the product.
	They also focus on \textit{situatedness} and \textit{temporality} of experience~\cite{Hassenzahl2006b}.
	
	\item \textit{Beyond task-related aspects of software use (aka. beyond instrumental)}:
	Models in this approach (e.g.~\cite{Hassenzahl2003a,Jordan2002a}) emphasize breaking down UX into a number of underlying elements.
	They argue that although UX is not fully predictable, it is to some extent \textit{shapeable} through the control of these underlying elements.
	In this approach, a positive UX is argued to be facilitated through  \textit{satisficing}\footnote{as opposed to satisfying functional requirements~\cite{Chung2000a}} the end users' non-task-related (aka. non-instrumental or hedonic) needs, as well as their task-related (aka. instrumental or pragmatic) needs.
	For instance, Hassenzahl~\cite{Hassenzahl2003a} views underlying UX as pragmatic (concerning performing tasks) and hedonic (concerning psychological well-being of the users) attributes.
	According to Hassenzahl, the user's perception of these attributes leads to some consequences: a judgment about the product's appeal (e.g., ``It is good/bad''), emotional consequences (e.g., pleasure, satisfaction) and behavioral consequences (e.g., increased time spent with the product). 
	He further emphasizes that these consequences (i.e., satisfaction, pleasure, appeal) are \textit{outcomes of experience}.
	
	\item  \textit{Emotion, mood and affect}: 
	Models in this approach (e.g.~\cite{Desmet2002a,Norman2004}) emphasize human emotions, and aim to understand  the role of affect \textit{``as an antecedent, a consequence and a mediator of technology use''} as Hassenzahl, and Tractinsky highlight~\cite{Hassenzahl2006b}.
	In this approach, a positive UX is argued to be facilitated through controlling and evoking positive emotions in users.
	
	\item \textit{Integrated experience}:
	Models in this approach (e.g.~\cite{Zimmermann2008a,Thuring2007a}) combine the two latter approaches, through integrating task-related and non-task-related aspects of experience, as well as emotional user reactions, to achieve an integrated user experience perspective.
	One example is the work of Th\"uring and Mahlke's~\cite{Thuring2007a} in which UX is divided into three main components of \textit{instrumental} (concerning usability and usefulness), \textit{non-instrumental} (concerning look and feel) and \textit{emotional reactions}.
	They emphasize that perception of users from the instrumental and non-instrumental qualities of a piece of software can lead to \textit{episodes of subjective feelings} (i.e., emotions).
	These repeatedly occurring episodes at the end shape the user's \textit{emotional experience}.
	\end{itemize}

Hassenzahl~\cite{Hassenzahl2010a} describes experience as \textit{``both unique but at the same time emerging from distinct elements and processes which are open to study and deliberate manipulation in an act of design.''}
Such an experience is, as emphasized by Hassenzahl, unique to the situation - time and context:
\textit{``Experience emerges from the integration of action, perception, motivation, and emotion, however, all being in a dialog with the world at a particular place and time.''}
Thus, whether we approach it holistically or not, UX is an emergent phenomenon~\cite{Hassenzahl2006b}.
	UX of a piece of software, among other things, emerges from underlying functionalities and quality characteristics, and the user's perception of them, in each certain situation.
\textit{Concrete} quality characteristics, therefore, can contribute to satisficing certain \textit{abstract} quality characteristics that designers find relevant to the software, e.g. in order to be trustworthy (abstract) the system provides a good overview of the functions available (concrete)~\cite{Hassenzahl2001a}.

The emergent nature of UX also highlights its \textit{temporality} meaning that the experience of a user with a piece of software can change over time.
As McCarthy and Wright~\cite{Wright2010a} emphasize:
\textit{``We are continuously connected emotionally into situations. Anticipation and expectation connect past experience to present and future experience.''}
This means that the overall experience is influenced by not only the interaction of a user with software, but also what they anticipate for this interaction, or remember from the interaction~\cite{ISO9241}.
Researchers therefore recommend taking the whole spectrum of interaction into account when studying the UX of a piece of software, in particular when evaluating it~\cite{Wright2005}.
This means studying the user's experience  \textit{before}, \textit{during} and \textit{after} the user's interaction: 
(i) \textit{anticipated experience}, what the user expects to experience~\cite{Wright2010a},
(ii) \textit{momentary experience}, what the user is experiencing during or immediately after interaction~\cite{Thuring2007a},
and (iii) \textit{lasting experience},what the user remembers of the experience~\cite{Hassenzahl2003a}.

All in all,  as Hassenzahl~\cite{Hassenzahl2010a} emphasizes, UX \textit{``may not be fully explainable and predictable from single underlying elements''} or `entirely reducible' to these elements; but it is still shapeable and controllable through them~\cite{Hassenzahl2010a}.
UX therefore can be manipulated through an act of design, i.e., designers' choices~\cite{Hassenzahl2010a}.
Considering the emergent nature of UX, current UX-related requirements work emphasizes identifying and refining abstract quality characteristics into functionalities and concrete quality characteristics in close collaboration with the end users' representatives~\cite{Hassenzahl2001a}.
In addition, current approaches to UX evaluation focus on both the end users' often narrative opinion of the system and  gathering statistically significant amount of data on users' reactions and perceptions in order to measure UX.
We should however note that researchers or practitioners still have not achieved a consensus about these UX measures, and their applicability in practice~\cite{Law2014b}.
Temporality and context dependency of UX also plays an important role in the evaluation approaches.
For instance, these approaches separate a user's perception of short-term and lasting experience.
An overview of various approaches to UX evaluation and measurement can be found in the work of Law et al.~\cite{Law2014b} or Mahlke~\cite{Mahlke2008a}.

Current UX models have all been developed outside the field of software development, i.e., not in computer science or Software Engineering (SE) literature, and can be argued to have had only a moderate effect on both.
Within these fields, researchers have responded to the growing importance of UX by advocating that ISO/IEC standards on \textit{software quality models} need to be extended to incorporate UX~\cite{Bevan2008a}.
This has to some extent been achieved in ISO/IEC~25010~\cite{ISO250102011} through introducing the Quality in Use (QiU) model.
This standard defines QiU as:
\textit{``the degree to which a product or system can be used by specific users to meet their needs to achieve specific goals with effectiveness, efficiency, freedom from risk and satisfaction in specific contexts of use.''}
QiU is similar to UX in that it also emphasizes users' personal (aka. non-task-related) needs and emotional reactions.
This model includes 'pleasure' (i.e., an emotional consequence of interacting with a piece of software) as a quality characteristic and defines it as:
\textit{``degree to which a user obtains pleasure from fulfilling their personal needs''}.

Developing more clear definitions and models of UX, however, is not by itself sufficient to facilitate its integration into software development processes.
Practitioners also need access to suitable tools, methods and techniques, and learn how to integrate them into the current development processes~\cite{Law2014a}.
Broadly speaking, the software community has adopted two approaches to address this need of practitioners.

The first approach has focused on developing and proposing new UX practices.
For instance, Doerr et al.~\cite{Doerr2007} propose a systematic guideline for measuring user satisfaction in early development phases.
They aim to assist practitioners in prioritizing requirements based on how these requirements contribute to a better UX in the software.
Kerkow~\cite{Kerkow2007b} and Nass~\cite{Nass2012a} propose an elicitation approach that facilitates finding a balance between the task-related and non-task-related requirements of software for a better UX.
Nass et al.~\cite{Nass2012a} argue for finding a balance between business and user goals, and present an integrated software development approach to address the challenges practitioners face in finding the right balance between the task-related and non-task-related aspects of software use.
Thew and Sutcliffe~\cite{Thew2011} present a technique that helps practitioners elicit emotions, values, and motivations; what they refer to as `soft requirements'.

The second approach directly or indirectly describes current challenges in UX integration, i.e., what are problems that practitioners face when doing UX work.
As mentioned earlier, studies in this approach can be divided into three categories, that we detail in the following sections.
None of these studies have analyzed the implications of similarities and differences between UX and other software quality characteristics, particularly usability.
Our study aims to complement the studies in the this approach by investigating UX challenges in a variety of software organizations with different development processes.
Also, in our analysis we pay certain attention to how identified challenges relate to the practice of software quality characteristics in general, and usability in particular.

\subsubsection*{UX challenges - specific aspects of UX work:}
The first group of studies that report UX challenges have focused on only some aspects of UX work: Measurability of UX (e.g.~\cite{Law2014b,Gerea2015a}), UX evaluation (e.g.~\cite{Alves2014a,Vermeeren2010a}), or understanding of UX (e.g.~\cite{Lallemand2014a})

For instance, Lallemand et al.~\cite{Lallemand2014a} investigated practitioners' understanding about the concept of UX, including their opinion on a number of UX definitions.
This was achieved through an international survey study with 758 participants including practitioners and researchers with both UX and non-UX-related backgrounds.
They provide an overview of how practitioners currently perceive the concept of UX and its importance in the practice  of software development.
They also discuss one main challenge in UX work: \textit{the low impact of UX research on industry}.
They stress that various research findings about UX (e.g., its relation to usability) is not still well received in practice.
They therefore emphasize to better integrate UX theories in industry, and educate students on UX research.

Law et al.~\cite{Law2014b} explore the basic question of whether UX constructs are measurable. 
Their data is based on semi-structured interviews with 11 UX professionals as well as 170 survey responses from the Human Computer Interaction (HCI) community.
Their paper also reports the results of a systematic review of 58 different papers from 2005 to 2012 on UX  measures and provides an overview of attempts to make UX measurable~\cite{Law2014b}. 
Law et al. describe how their interviewees expressed skepticism and ambivalence towards specific UX measures even if attitudes were more positive overall.
Respondents considered most UX constructs to be measurable.
Nevertheless, Law et al. note that practitioners show opposing views on whether UX can or should be divided into composing elements, or whether it needs to be considered or measured as a whole.
Results from their interviews show three categories of challenges when it comes to the interplay between UX evaluation and software development:
 (i) theoretical (measuring UX holistically or in elements, and conceptualizing long-lasting versus momentary experience), (ii) methodological (differing preferences for quantitative versus qualitative data by design- and engineering-oriented stakeholders), and (iii) practical (lack of knowledge and competence for interpreting measurement outcomes).

The survey by Law et al. is duplicated in the context of the Latin American software development industry~\cite{Gerea2015a} through 40 (out of 112 sent out) survey responses.
According to the survey results, practical aspects such as cost and time play a more important role in whether or not practitioners measure UX in Latin America.
Gerea et al.~\cite{Gerea2015a} discuss this result as being quite different from the original survey in which these two issues are not included as formal practical issues.
Other challenges reported by Gerea et al. include: \textit{limited access to the end users} and \textit{lack of knowledge and experience in UX measurement}.

Through 97 survey responses by software development practitioners with different backgrounds,  Alves et al.~\cite{Alves2014a}  investigated how \textit{UX evaluation} is performed in practice (i.e., by whom, in what phases of software development, and using what tools and methods).
They report a number of challenges concerning UX evaluation.
For instance they highlight the challenge of \textit{low involvement of users}.
According to their data, in around 50\% of the cases UX evaluation is performed without involving the end users, and often evaluators `assume' what the perception of users will be, rather than actually involving them.
This challenge is attributed to the generic issue of \textit{limited resources} (time, effort, and competence).
Another challenges concerns \textit{low involvement of UX experts}.
Alves et al. report that sometimes evaluations are performed by software developers that do not necessarily have the required competence.
Another  challenge according to them concerns \textit{tools and methods}, and that often selection of tools and methods is motivated by cost rather than suitability for the project.
In their study, Alves el al. used a list of evaluation tools and methods that are mainly usability specific (e.g., cognitive walk through, task analysis), or generic (e.g., observation, interview, think aloud).
The list lacks tools and methods that are proposed for evaluating other aspects of experience for instance the non-task-related aspect or user emotional reactions (e.g., Attrakdiff questionnaire~\cite{Burmester2015}\footnote{see \url{http://www.allaboutux.org} for more UX specific tools and methods}).
This can introduce a risk to the data because practitioners might have preferred the use of a generic method such as a questionnaire for evaluating UX, without necessarily acknowledging that such generic methods may purely produce usability-related data if not used with specific attention to UX-related concepts, such as the non-tasks-related aspect of use~\cite{Hassenzahl2010}.

Similarly, Vermeeren et al.~\cite{Vermeeren2010a} performed a combination of empirical research methods (including an online survey) to investigate the current state of UX evaluation, and identify what evaluation methods are more often used by practitioners.
They identified 96 methods and provide a good overview of such methods and their use in practice.
They also report a number of challenges practitioners face when applying these methods and argue that some still need to be further improved and developed for better use in practice.
These challenges include: \textit{lack of suitable methods for evaluating UX in earlier phases},\textit{ lack of methods that explicitly study the experiences of groups of individuals}, \textit{lack of methods for evaluating UX in the period before actual use} (i.e., anticipated use), \textit{limited practicality of some current methods}: e.g., because they need special competences, are time consuming, or analyzing their results is difficult.

\subsubsection*{UX challenges - UX work in agile projects:}
The second group of studies that report UX challenges have focused on UX work in agile projects.
For instance, through document analysis and interviews with seven practitioners in one case company, Isumursu et al.~\cite{Isomursu2012a} report that there are still \textit{uncertainties about organizing UX-related roles and responsibilities} in agile projects.
This in turn can lead to \textit{communication problems between UX and non-UX practitioners} involved in the project, and also \textit{difficulties in integrating UX practices with other agile design and development activities}.
Isumursu et al. also report that \textit{visibility of UX targets} (set to drive UX-related decisions) decreases throughout projects. 

Another study on UX work in agile projects is performed by Larusdottir et al.~\cite{Larusdottir2012a}.
They investigated integrating UX related activities into Scrum projects through two in-depth interviews with two UX experts.
They report that these experts' biggest challenge in Scrum projects concerns \textit{losing the big picture of UX design.}

To investigate strengths and weaknesses of Kanban and Scrum concerning UX work, Law et al.~\cite{Law2015a} performed a conceptual analysis of these approaches, followed by interviews with 10 practitioners and 73 survey responses from software development practitioners.
The authors conclude that practitioners find it very challenging to fully integrate UX work with these two approaches.
Law et al. relate this to \textit{the fundamental differences between the philosophies, methodologies, and practices of these two approaches compared to UX work} and that Kanban and Scrum are developer- and customer-oriented, not user-oriented.

\subsubsection*{UX challenges - mixed with usability:}
The third group of publications on UX challenges have studied both UX and usability without differentiating between them (e.g., in agile development processes~\cite{Cajander2013a,Ovad2015a,Kuusinen2015a}, or in software development in general~\cite{Ardito2014a,Ardito2014b,Lanzilotti2015a}).
For instance, Ardito et al.~\cite{Ardito2014b} investigated the current practice of UX and usability through a four stage empirical study including an online survey (with 36 participants from Italy, and 39 participants from Denmark), interviews with four practitioners from four different companies, a focus group, and an exploratory study.
Although the authors emphasize both concepts of UX and usability, their survey includes only questions about usability evaluation.
Similarly, their interview study focuses on practitioners' understanding and experience of usability evaluation.
The results of their survey show that still several software development companies do not perform any usability evaluation.
Practitioners often relate this to that such \textit{evaluations are highly resource demanding}, \textit{access to the end users is limited}, and \textit{there is a lack of suitable tools and methods} to support performing evaluations.
One other explanation for limited usability evaluations in the participating companies according to Ardito et al. is that \textit{UX and usability requirements often are not included in the requirements documents}.

The study of Ardito et al.~\cite{Ardito2014b} is complemented by Lanzilotti et al.~\cite{Lanzilotti2015a} through 54 survey responses by software development practitioners, and the analysis of 44 requirements documents. 
Lanzilotti et al.  aimed to understand the current state of UX, and usability requirements in particular, in \textit{call for proposals}.
According to their findings, such requirements are either not included in these documents, or are included vaguely.
Similar to Ardito et al.~\cite{Ardito2014b},  Lanzilotti et al. conclude that a \textit{lack of formal requirements for UX and usability} is one main reason why UX and usability practices are ignored in projects, or lack assigned resources.
The authors do not however differentiate between UX and usability requirements, and seem to use the terms interchangeably, as also goes for the study of Ardito et al.~\cite{Ardito2014b}.

Ovad and Larsen~\cite{Ovad2015a} performed two interview studies in 2013 and 2015 in order to investigate the maturity of UX practices in agile development environments over time in eight participating companies.
This study reports a number of challenges practitioners faced in their work in 2013 and discusses how these challenges were addressed over a two-year period.
These challenges include \textit{lack of practitioners' knowledge on UX},\textit{ ignoring UX in development projects} (e.g., because of \textit{low management support} or that \textit{UX practices are not considered elements  of the process}),
\textit{lack of a defined UX process}, \textit{failing to assign the responsibility of UX}, and \textit{assigning limited resources to UX work}.
The authors discuss how, except for the last challenge, other challenges were at least to some extent addressed in those companies by 2015.

	Cajander et al.~\cite{Cajander2013a} interviewed 21 practitioners regarding challenges they face in integrating \textit{user perspective} in Scrum projects (which can influence both usability and UX).
	They conclude that there is \textit{no clear picture of the responsibility for usability}, and that \textit{usability goals are often unclear in projects}.
	According to their data, \textit{user involvement and design feedback is often ad-hoc} and usability and UX practitioners often \textit{lack suitable methods} to support their work in Scrum projects.

	Kuusinen~\cite{Kuusinen2015a} investigated the task allocation between UX professionals and other team members in agile projects.
	Although they use the term UX they express that they have merely focused on usability and functionality since that is how industry currently approaches UX.
	Through a longitudinal multiple case study, Kuusinen shows that often \textit{UX collaborations focus on Graphical User Interface (GUI) design and other aspects of UX work are downplayed}.

\section{Methods}
\label{sec:method}
We have conducted an explorative, qualitative study~\cite{Braun2006a} to investigate the challenges involved in integrating UX practices into software development processes.
Below, we detail our research approach by describing the different companies where we conducted interviews, how the data was collected, and our approach to data analysis.

\begin{table}
\begin{scriptsize}

\begin{tabular}{|p{0.2\textwidth}|p{0.1\textwidth}|p{0.64\textwidth}|}
\hline
Company Type & Number of Employees & Interviewee(s) \\
\hline
(A) IT services, consultancy \& outsourcing & 	>130,000	& UX designer
\\
\hline
(B) Defense  \& security	& \~14,000	&
Technical project manager,
Technical project manager,
Product manager,
Technical project manager,
Technical project manager,
Developer
\\
\hline
(C) Wireless solutions \& network testing	& \~1,800	 &Interaction designer
\\
\hline
(D) Systems \& software development company 	& \~ 120	& Developer
\\
\hline
(E) User Experience consultancy \& training	  & \~ 50	 & UX designer
\\
\hline
(F) IT \& management consultancy	& > 1,300 & 	Art director,  Interaction designer, Developer, Project manager, Art director
\\
\hline
(G) User Experience \& Usability consultancy	 & 34	& UX designer
\\
\hline
(H) Telecommunication	& \~ 114,000	& UX designer, UX designer
\\
\hline
\end{tabular}
\caption{Overview of the companies and interviewees that participated in the study}
\label{tbl:sites}
\end{scriptsize}
\end{table}

\subsection{Research Sites}
We selected a variety of companies with different characteristics for our study in order to improve the generalizability of our findings~\cite{Runeson2008a}.
Table \ref{tbl:sites} shows an overview of these companies and their main characteristics and labels them (A-H) for easier reference later in the paper.
The companies span various domains (company type) and vary in size (number of employees). The table also details the roles of the interviewees at each company.
The first two companies are active only in Sweden, and the rest are internationally active.
We approached both consultancy and product development companies in order to cover both perspectives.
Both A and E are well-known consultancy companies in Sweden, while B, C, D, and H are well-known product development companies in Sweden.
Throughout the study, we were introduced to other companies by our interviewees, from which we also included a number of interviewees (F and G).
Only one of the companies (E) was previously known to the authors based on previous research collaborations.

\subsection{Data Collection}
At each company we asked for interviewees with specific roles such as practitioners with technical or design backgrounds.
The aim of the interviews was presented to our main contacts in each company to make sure that the selected interviewees were suitable for our research.
We had the option to ask for more interviewees, but since the study was explorative, after \intervieweesnum interviews we were confident that we had covered the major challenges from a sufficiently broad range of perspectives.
The selected  practitioners (See Table~\ref{tbl:sites}) represent technical (e.g., developers), design (e.g., interaction designers), and management roles.

We conducted semi-structured interviews~\cite{Flick2009} to collect more of the interviewees' viewpoints, which was important to the explorative nature of our study.
We prepared an interview guide with a set of pre-designed questions based on the knowledge gained from literature (Appendix A).
In each interview, questions were rephrased, added, or skipped based on the interviewee's background and responses.
Each interview covered all phases of software development processes, the activities performed in each phase, and the tools and methods applied.
Ample time was spent exploring and discovering current approaches to UX and challenges that were not known beforehand.
Thirteen of the interviews were conducted face-to-face, and four via video or telephone conference.
Each interview lasted between 30-60 minutes, and was recorded and transcribed.
The interviews were all performed in the spring of 2012.

\subsection{Data Analysis}
We analyzed the data by applying a combination of inductive and deductive approaches, generally known as \textit{thematic analysis}.
Thematic analysis is defined as \textit{``a method for identifying, analyzing and reporting patterns (themes) within data.''}~\cite{Braun2006a}.
In order to do so, we \textit{segmented} the interview transcriptions into meaningful paragraphs or sentences in a way that each of these segments presented one concept; we then \textit{coded} these segments~\cite{Braun2006a}.
In the coding process, the `key points' of each segment were first noted, then codes that were most suitable to these key points were identified.
Based on a pilot coding, we gradually and iteratively generated an improved coding guide for the rest of the coding and analysis process.
The initial main codes that were used in this study include definition, challenges, solutions, tools and methods, evaluations, requirements, UX versus usability, and activities.
These codes reflect a broader scope than merely challenges because we were interested to learn the identified challenges in context.
Therefore, when applicable, during the interviews, follow-up questions were asked regarding the mentioned challenges.
These questions aimed to gather more data on attempted solutions, or previous experiences with usability or other quality characteristics.
We then further analyzed the segments that were coded as `challenges' in more detail.
The initial list of codes were generated based on our knowledge and experience in the field and after we became more familiarized with the interview data~\cite{Braun2006a}.
When the interviewees used different terminologies, or had limited knowledge concerning UX or usability, we mapped their statements to the relevant concepts based on the definitions of the concepts in literature.

To make our analysis process concrete, we give an example of how we performed coding and identified challenges.
The interview segment:
\textit{``There is a bit of confusion in the field and in the company as well, what's the difference? design is design''} (A-1) resulted in identifying key points:  (definition of UX is not clear, practitioners do not know what UX exactly means), and was coded with `challenge', and `definition'.
This segment was further analyzed because it was coded as `challenge'.
Together with other segments that related to understanding and definition of UX, this segment then resulted in creating the challenge \textit{`lack of consensus on definition and construct of UX'}.

Our analysis resulted in \challengesnum challenges.
In reporting these challenges, we have not separated issues such as \textit{lack of resources}, or \textit{low management support} that are too generic to provide new insights to the community.
We instead have discussed these issues in the context of other challenges where applicable.
There is a clear multifaceted relation between the identified challenges.
Some, for instance, could be seen as causes or at least underlying factors in relation to other challenges.
Some are more fundamental and concern the view, attitude and knowledge of stakeholders while some are more tactical and/or practical.
Still, in this paper, the challenges are presented as one linear list rather than as an interconnected `web' of challenges. 
This is mainly because our data does not provide sufficient evidence to support the relations between identified challenges. Future work should consider studying connections between challenges in more detail.

\subsection{Threats to Validity}
Threats to validity are outlined and discussed based on the classification by Runeson and H\"{o}st~\cite{Runeson2008a}:
\begin{itemize}

\item Selection process of subjects for interviews can cause a threat to \textit{construct validity}.
Selection bias is always present when subjects are not fully randomly sampled.
However, here the subjects were selected based on their role, experience and availability so there is little more we could do to alleviate this threat.
The presence of a researcher may influence the  behavior and response of the subjects.
This threat was alleviated somewhat by the guarantee of confidentiality of the data but is an inherent aspect of the research method used.

\item In any empirical study, incorrect data is a threat to \textit{internal validity}.
In case of the interviews, taking records in form of audio, which was then transcribed, mitigated this threat. The authors also analyzed the material in several rounds of independent as well as joint sessions to gradually reach consensus on the intended meaning of the respondents.
We also shared the results of our analysis with the interviewees to validate and confirm the findings.

\item \textit{External validity} concerns the ability to generalize the results beyond the actual study.
Since the interviews are just a sample from a potentially very large population, they should be interpreted with some caution. Still we sampled a number of different organizations in different industrial domains to decrease the effect of this threat. 
However, qualitative studies rarely attempt to generalize beyond the actual setting and are more concerned with explaining and understanding the phenomena under study.
Another concern is that our data gathering was performed in the spring of 2012. Therefore, our data may not reflect today's UX state-of-practice in these organizations. However, the data still is valid when interpreted in its own time frame.
Also, to minimize the effect of time span on our analysis, we have included recent studies published since 2012 when analyzing the data and discussing the results. 

\item Another threat concerns \textit{reliability}, to what extent the data and analysis are dependent on the specific researchers.
In order to minimize this threat in analysis, the three authors individually and independently conducted a pilot coding of these segments using an initial coding guide as explained above.
The outcomes of the pilot coding were discussed in several sessions with all three authors, and the differences in coding were analyzed and resolved.
Also, we had carefully designed the interviews before running them, and the coding process before analyzing the data.
\end{itemize}

\begin{table}
	\centering
\begin{scriptsize}
\begin{tabular}{|p{0.75\textwidth}| }
\hline
\textbf{The identified challenges}
 \\
 \hline
 C\rchl{ch:kn}.~Lack of consensus on definition and construct of UX
 \\
 C\rchl{ch:value}~Lack of consensus on the value of UX
 \\
C\rchl{ch:models}.~Low industrial impact of UX models, tools, and methods
\\
C\rchl{ch:techfocus}.~More focus on objectively measurable aspects
\\
C\rchl{ch:re}.~Difficulties in engineering UX-related requirements 
\\
C\rchl{ch:uxeval}.~More focus on testing functionalities and usability than UX evaluation
\\
C\rchl{ch:lackcomp}.~Lack of consensus on UX-related competences and responsibilities
\\
C\rchl{ch:late}.~Late focus on UX in projects
\\
C\rchl{ch:communication}.~Communication and collaboration gap between UX and non-UX practitioners
 \\
C\rchl{ch:customer}.~Customers resistance to the cost of UX practices
\\
C\rchl{ch:user}.~Low user involvement
\\
 \hline
\end{tabular}
\caption{We identified \challengesnum challenges through interviewees with \intervieweesnum practitioners in eight different software development companies}
\label{tbl:chl}
\end{scriptsize}
\end{table}

\section{Results}

This section presents the challenges discovered during the analysis of the interview data.
As mentioned earlier, there is clearly a multifaceted and complex set of relations between these challenges.
Some for instance could be seen as symptoms (e.g., C\ref{ch:techfocus} and C\ref{ch:late}) and some as explanations to other challenges (e.g., C\ref{ch:kn} and C\ref{ch:models} ).
Some are more fundamental and concern the views, attitude and knowledge of stakeholders (e.g., C\ref{ch:kn}, and C\ref{ch:customer}.) while some are more tactical (e.g., C\ref{ch:re}, and C\ref{ch:uxeval}.).
However, we do not have enough evidence in our data to support these detailed classifications and relations because our aim was a broad exploration of challenges, not to validate specific relations and connections.
Thus we have decided to report these challenges in a list here.
Section~\ref{sec:summary}) then discusses a number of possible interrelation among these challenges to the extent observable in our data.

\subsection{The Identified Challenges}
\label{sec:challenges}

We have structured our interview results in a list of \challengesnum challenges, summarized in Table~\ref{tbl:chl}.
These challenges are presented in the following in more details, and supported by interviewees quotations.

\subsection*{C\ref{ch:kn}.~ Lack of consensus on definition and construct of UX}

Our data shows that practitioners' understandings of the concept of UX differ and is, in some cases, even inconsistent and contradictory.
According to the respondents, UX is a new concept and there is still a lack of general agreement on its meaning in the field in general, and among practitioners within organizations in particular (A-1).
In some  practitioners' view, UX is the same as, or is seen as, usability or Interaction design (IxD).
This is while these two concepts only concern details of the interaction, and/or the Graphical User Interface (GUI) design as the UX practitioners stressed.
They also emphasized the reason for focusing more on usability and IxD is their relative simplicity:
\textit{``I think discussions at large when it comes to UX design at common ground is still about IxD and usability.  Usability is easy to talk about and everybody understands it.''} (A-1).

One of the interviewees referred to UX as `just another buzz word' (E-1).
In her view, UX contains the same concepts that have been around for a long time under other names such as usability and `emotional design'.
On the other hand, some practitioners explicitly differentiated between UX from Usability (e.g. H-1) or IxD (e.g. F-1).
One of the participants emphasized that usability is `a minor subset of UX' and added:
 \textit{``I've never, never called myself a usability expert, and I would never do that.''} (H-1).
Some of the participants emphasized that UX goes beyond `cognitive aspects of design' (e.g. E-1), the main focus of usability.
Similarly, another practitioner stressed that UX is not about `IxD' and has a much broader perspective:
\textit{``[IxD] is just the end results, we do not call ourselves interaction designers, because that is only 10\% of our work but that is an important element because that is the most visual part of our work .''} (F-1).
Similarly, another practitioner referred to UX as:
\textit{``a wholeness with the emotional, social, economical, functional, and technical parts.''} (A-4).
Another practitioner described UX as:
\textit{``pretty much everything that affects a user's interaction with a product.''} (H-2).
He further emphasized UX is `the whole package' and usability is only one part of it.
Some practitioners' related UX to the `why' behind the functional requirements and the software in general.
For instance, one practitioner described this through three main questions of `what' to build, `how' to build it, and `why' to build it (A-4).
Some practitioners related UX to the graphical user interface (GUI) by stating that it is about \textit{``the cool things, the new things, the flashy things.''} (A-3).
In their view, UX is mainly about \textit{emotions} and \textit{aesthetics} therefore not applicable to all types of applications, for instance `productivity applications'.
 
In general, the practitioners with more technical backgrounds and roles showed less knowledge about UX.
Their knowledge was often limited to cognitive aspects of design, i.e., usability.
For instance, one of the technical managers stated:
\textit{`` [our customers] talk about increased workload. That is a negative thing.
	I don't know if that qualifies as UX.''}  (B-2)
A number of practitioners stated that UX is a `broad' and `holistic' concept covering not only the user perspective, but also the business perspective (e.g. H-1).
While the latter looks into how the design contributes to achieving business goals, the former assures supporting the end users' goals.
The user perspective includes aspects such as `emotions', `values' and `preferences'.

The participants also discussed that customers' limited knowledge about UX is a challenge.
For instance, they stated that customers who have heard about UX can be too ambitious regarding emotional and non-task-related user needs.
Customers' limited knowledge also means that they often specify the related requirements vaguely and using inconsistent and subjective terminology.
They often indicate a need for quality characteristics such as  `cool', `fun', or `high-tech' mostly because they are affected by such `buzz words'.
They often neither have knowledge about what these terms actually mean, nor find them specifically relevant for their products.
Therefore, practitioners emphasized that  to prevent misunderstandings, these UX-related requirements should, early on, be refined to more `concrete' requirements, and specified in a measurable way (e.g. B-3).
Regarding this one of the interviewees said:
\textit{``usually they say `we want something like that app', `we want it to be cool and high tech'. Then you have to initiate a dialog to find out what  they mean for this particular customer.''} (A-1)

\subsection*{C\ref{ch:value}.~Lack of consensus on the value of UX}
Generally speaking, our data shows that various stakeholders still have different views on whether UX is important or not.
Nevertheless, according to several of the interviewees, an important motivation behind UX is the growing general importance of software in recent decades.
Interactions users had with earlier software were limited to command-line interactions in software built to support existing manual work.
This has now transformed into a multitude of interaction styles (mouse, touch, etc.) and applications.
Software is now a large part of all aspects of most human life, and users are exposed to a huge variety of it.
This exposure to various software systems affects the experience and expectations of users.
Regarding this, one practitioner said:
\textit{``Users are meeting a lot of good things, and they are expecting good things all the time.''} (E-1).

According to practitioners, various businesses are learning from successful products in the market.
This has inspired not only market-driven but also business-to-business software projects.
This is evidenced by one of the interviewees saying:
\textit{``A lot of business-to-business applications are being informed by business-to-consumer web apps, software apps.''} (G-1).
In particular, in cases where a product has competitors, the motivation to improve UX increases.
One of the practitioners argued that UX is nowadays a `differentiator':
\textit{``Today I think that many companies do usable products, in order to distinguish a brand or a product we need to add an extra level to the product so that is really what I call UX. We need to take more things into account, emotions, and it needs to look great, and it is not only about being usable.''} (A-2).
In market-driven products, branding, emotional concerns, and relations with end users are important.
Therefore, these business are often more concerned about UX.
Another interviewee argued that in market-driven software, in particular game development, UX is `part of the common practice' (A-1) while this is not the case in business-to-business software. 
He emphasized this approach of market-driven projects should be `transferred to other projects' as well.

Some of the practitioners emphasized that UX is an important software quality characteristic that needs to be taken into account more in projects.
But in some functionality-focused organizations UX is still considered something `on the side' and not a core concept or value.
According to these practitioners, when a product is more `technical', the business units often focus on functionality, and UX becomes less important.
As one of the interviewees emphasized, in this case the business unit is not `that concerned with the look and feel' (A-4).
The practitioners were generally positive that more and more organizations even the technology-focused ones are learning about UX and the importance of taking it into account in their products (e.g. H-2).
For instance, one practitioner said that in their company, business units are nowadays showing less resistance towards UX and the importance of UX `starts to be visible' (H-2) to these units.

In practitioners' view, often customers are less aware of UX and its value, and in particular how it differs from usability and IxD.
In addition, they emphasized that preferences, values, and motivations of upper management involved in strategic decision making is, according to some practitioners, another reason why some organizations might undervalue UX, and show a less positive attitude towards it.
Regarding this, one practitioner stated:
\textit{``I think it's sometimes just the reason people go into business in the first place. \dots the people who are in it because they think that their product or service solves a real problem, they generally care about it more.''} (G-1).
He further discussed how individuals, in particular higher management, play an important role in whether UX is a priority in an organization or not.
Regarding low support from other stakeholders, and in particular managers, one of the interviewees stated that the inputs from their UX group in `research and development' is ignored by business units because some of the ideas are ahead of their time:
\textit{``business units are occupied with their very close, near time results so they look at what can they sell now.''} (H-2).
He further highlighted that some of these previously rejected UX ideas are now being incorporated in the products and are getting support from management because competitors are now implementing similar ideas.

Overall there is a very varied view, and a general lack of consensus, on the value of UX both between and within the investigated organizations.
In addition, practitioners stated that customers also differ in their levels of awareness and how they value UX.
The common pattern is that customers often down-prioritize UX and are either not, or less willing to pay for it.

\subsection*{C\ref{ch:models}.~Low industrial impact of UX models, tools and methods}
While C\ref{ch:kn} concerned the lack of consensus on the definition and construct of UX itself, here in C\ref{ch:models} we report on difficulties in how this understanding is gained or put into practice.
This challenge is also in direct relation to C\ref{ch:re} and C\ref{ch:uxeval} where our findings regarding testing or requirements activities are presented in more detail.
The reason for separating this challenge from C\ref{ch:re} and C\ref{ch:uxeval}  is that we did not want to duplicate these difficulties in two different sections.
In addition, by using a separate section we put more emphasis on the important roles of models, tools and methods in the practice of UX, as our data reveals. 

We observed that often practitioners' knowledge regarding UX is more based on experience and work with similar concepts, not on any specific UX models or theories.
This knowledge is  sometimes shaped through their education on `usability' and IxD, and has changed over the years to be focused more towards UX.
For instance, one of the practitioners stated:
\textit{``When I first started in 90s, there was no such thing as UX or IxD, actually. 
	[\dots then] it has been IxD and then sort of merged into experience, UX design.''} (H-1).

Very few of the interviewees were familiar with currently influential approaches to UX and corresponding models, even those interviewees that demonstrated a relatively good understanding of UX.
For instance some interviewees said the way they look into users' emotions, values etc. is not `in an academic way' (e.g. A-4)  or `by the book' (e.g. A-2) but based on experience.
Still, practitioners generally showed a positive attitude towards applying new models, tools, methods and techniques to their work:
\textit{``we are lacking this, so this would be really nice to have more research results that we could apply.''} (A-2).
Still, according to the practitioners we spoke to, some organizations resist introducing new models, tools, methods, or techniques.
This means that in these organizations, practitioners can only rely on traditional interview and observation techniques when performing UX practices (A-2).
C\ref{ch:re} and C\ref{ch:uxeval} further detail more issues concerning limited access to tools, methods, and techniques for supporting UX requirements and their evaluation.

The interviewees referred to two models they use in their UX work: `emotional design' by Donald Norman~\cite{Norman2004} and Maslow's hierarchy~\cite{Maslow1943a} (F-1).
The latter is used as an inspiration when eliciting user needs.
The interviewees mentioned such models can help to create a methodology to work with UX and `build the right things in the right order' (F-1).
But respondents with this type of experience were a clear minority.

\subsection*{C\ref{ch:techfocus}.~More focus on objectively measurable aspects}

A group of practitioners emphasized that the software development and engineering community has traditionally had much greater focus on software functionality than quality characteristics.
We can explain the identified challenges concerning UX-related requirements (C\ref{ch:re}) and testing (C\ref{ch:uxeval}) through this lopsided emphasis on functionalities and actual quality characteristics.  
Since C\ref{ch:techfocus} relates to both requirement and evaluation, it is presented in a separate section to prevent duplicating the issues in two places.

In their view, functionalities of a piece of software are important in achieving a good UX but too much focus on them can often lead to ignoring UX (e.g. H-2).
One practitioner stressed the relation between functionality and UX as follows:
\textit{``The quality of experience is really depending to some extent on how the functional requirements are met, but also actually on what the functional requirements are, also just the amount of them.''} (H-1).
The interviewees emphasized that satisfying functional requirements does not necessarily mean that the correct or valuable functionality is included in the software.
This is evidenced by one interviewee saying:
\textit{``what do you know when you have signed [the technical specification]? Do you know that it is a good solution? No! you only know that it meets the functional requirements and to me it is silly!''} (E-1).

In addition, according to the participants, the software community still often focuses more on `actual' qualities of software than its`perceived' qualities.
While the former concerns objectively measurable quality characteristics, the latter concerns how users subjectively perceive these qualities.
For instance, users may perceive a five milliseconds response time (i.e., actual performance) as fast or slow (i.e., perceived performance).
Regarding the role of perceived qualities in experience of users an interviewee stated:
\textit{``sometimes the perception of time is more important than the actual time, and these are the things you should pinpoint [to the stakeholders]''} (E-1).

\subsection*{C\ref{ch:re}.~Difficulties in engineering UX-related requirements}
According to the practitioners, in many cases, the non-task-related needs of users or their emotions are still either neglected or treated only informally.
In one organization, emotional design goals are often only documented and communicated in the form of a `post-it note on the wall', as a reminder.
The practitioners highlighted that it is still an open problem as to how to map these types of needs to measurable requirements.
For instance, one of the interviewees stated:
\textit{``I would say the emotional part of this is very very rarely formally put into words.''} (A-1).
Practitioners argued that these needs are hard to elicit, define, communicate, and agree upon.
Stakeholders in general have less knowledge about this type of needs, organizations still lack the related competencies and have only limited access to tools, methods and techniques to deal with it (e.g. B-3).

Emphasizing the challenges concerning UX-related requirements, one interviewee stated:
\textit{``Functional requirements are easy to create, to merge into a design; more emotional things are more difficult.''} (F-1).
Similarly, one interviewee stated that features are `easier to define' (H-1).
Besides lack of knowledge and understanding, practitioners related these problems to a lack of competence in dealing with UX-related requirements within their organizations (e.g. A-3).
Regarding this, one practitioner highlighted:
\textit{``features are what most project managers, most managers can understand.  You can count them, you can map them to customers, customer dialog for instance, and so forth, and you can compare your amount of features with the competitors.''} (H-1).
Similarly, another practitioner stated:
\textit{``I think it is largely a competence thing. Doing emotional aspects of design is quite a new concept, I have only heard about it in the last year, or last two years maybe, so I do not think that knowledge has really reached the industry yet.''} (A-3).
In particular, when the software product is more innovative these problems are compounded.
This is evidenced by an interviewee saying:
\textit{``This is a kind of project where nobody really can tell how this should be, what it should be like, nobody has done it before, there are no standards to refer to \dots	who can specify those [UX-related] requirements? You need to have a certain quality  but what is the level of that quality? We haven't really found what that level is.''} (B-3).

Some practitioners highlighted the challenge of finding a balance between UX-related needs, business goals, and technological constraints.
Regarding the importance of finding a balance between emotional and business needs, one of the interviewees stated:
\textit{``you can spend a lot of time, thinking about people's emotions and so on, but if you are going to succeed you have to look at [business perspective].''} (E-1).

In general, UX practitioners argued that UX-related requirements should be elicited before refining functional requirements.
For instance, one interviewee stated:
\textit{``First you have to define the business requirements, the user requirements, [where] you have the IxD; and then you can define the FRs.''} (E-1).
Another practitioner said:
\textit{``I think that the functional requirements should come as a result of a dialog between different types of domains such as user experience, business, and technology.''} (H-1).
Nevertheless, according to these practitioners, such an approach is not common in practice.
This is expected since more technical roles emphasized that the requirement process should start by eliciting and refining functional requirements then quality requirements.

A practitioner stated that they use `persona' to `informally' document UX-related requirements:
\textit{``We'd specify this more in the persona descriptions; for example that this persona needs to, or wants to experience some kind of things.
In the wireframes, we might  specify an animation for example that it should feel smooth or something like that.''} (A-2).
They,however, further emphasized that this approach is not optimal and there is  a need for more formal approaches to deal with such requirements:
\textit{``it should be good if we could formalize it a little bit more I think.''} (A-2).

The practitioners  generally agreed that to communicate UX-related requirements, they require forms of requirements other than textual (e.g., sketches, wireframes).
They emphasized `concrete and tangible' form of requirements facilitate communicating the `fluffy' requirements.
Regarding this an interviewee told us:
\textit{``we create `mood boards' where you take an image-driven approach to the look and feel, and we use references of course, like `that app has a good flow in it', and `that app has a good feeling to it'.''} (A-1).
More technical practitioners also seemed to have a similar approach to UX-related requirements:
\textit{``If the customer said that they want it to `look nice', then you have to make the graphical design first and then they can say that `hey! this looks nice' and then you have taken care of that requirement.''} (A-3).

Nevertheless, only in one organization were these non-textual UX-related requirements (e.g., wireframes) traceable to business requirements (F-1).
Regarding this a practitioner stated:
\textit{``requirement list is not important for the business at the end, they wanna see the wireframes. So at the end we show the wireframes with all kinds of numbers and those numbers are linked to the excel sheets of the requirements, their descriptions and how they are linked with the CPR and business case.''} (F-1).

Problems with engineering UX-related requirements can be one main explanation for difficulties in evaluating UX (C\ref{ch:uxeval}).

\subsection*{C\ref{ch:uxeval}.~More focus on testing functionalities and usability than UX evaluation}
Our data shows that in general, the main focus of our selected organizations is on testing functionality.
We observed that practitioners with technical backgrounds are often less familiar with how their organizations handle UX or even usability testing.
They also showed limited knowledge as to why such evaluations can be useful to the success of the software (e.g., B-2, B-4).

We observed that, generally speaking, the practice of UX evaluation is still immature in many organizations.
In projects with limited time or budget, UX evaluation is either non-existent or rare compared to other testing activities:
\textit{``we have done much functional testing of course, system tests etc., but end user testing we have not performed much I'd say.''} (A-2).
In some organizations UX evaluation is basically replaced by usability testing.
Regarding this, one interviewee stated:
\textit{``I think user tests tends to be more focused on pure usability. I guess it's when you're releasing the product into the wild, that's when you start to get maybe the most valuable feedback or the most truthful ones.''} (A-4).
According to these practitioners, usability testing is not enough to evaluate the whole UX of the software.
Regarding this, a UX practitioner stated:
\textit{``when you evaluate usability, it's when you go into the nitty gritty details, and try to look at more efficiency within the user interface. My personal view is that that is not that relevant. I mean it's relevant, but not in what we do [i.e., UX].''} (H-2).
To compensate limited formal UX evaluations, some organizations gather informal qualitative user feedback after release, for instance through comments in the App Store or on social media (e.g. A-2).
Some organizations perform user surveys that are not necessarily designed for evaluating UX (e.g. A-1).

According to the practitioners we spoke with, UX evaluation is limited because it is more difficult compared to evaluating usability or other quality characteristics.
Practitioners gave various explanations for this difficulty.

Some practitioners related this difficulty to the fact that UX involves emotions, and non-task-related user needs, and that limited tools and methods exists to support these aspects.
In their view, emotion can be even impossible to measure using current quantitative approaches.
One of the practitioners stated:
\textit{``it is rather difficult to measure emotions and more softer issues I think. \dots
really getting the correct feeling that the user has. Because they will try to explain it but it perhaps is not the real emotion that we catch at the end. It would be nice to have some methods or approaches to extract this kind of information from the users.''} (A-2).
Similarly, one interviewee said:
\textit{``some goals are more difficult to measure than others, e.g., if this is a feeling thing: `I should be very well informed', but mostly you can measure [them] in the usage test through observations and interviews''} (E-1).
This practitioner further emphasized that although they can specify quantitative measures for UX-related requirements (e.g., ``10 out of 10 users should succeed, and they should be content''), they still need to observe users in order to gain better understanding of the experience:
\textit{``can the users perform the tasks? how do they perform the tasks? how do they feel afterwards? are they content?''} (E-1).

Some practitioners related the difficulty in UX evaluation to the holistic nature of UX.
Discussing cases where practitioners take a holistic approach to UX, one interviewee stated:
\textit{``how would you measure that sort of holistic experience throughout the process of designing it? 
Because, of course you cannot [implement or design] everything at the same time, and you know there are so many dependencies.
How do you straighten those out and how do you understand what you're measuring and not measuring?''} (H-1).
This  interviewee further emphasized that the broader scope of UX negatively impacts evaluation:
\textit{``That's a problem, of course, because UX is much broader in scope, and if you have a wider scope on it, then you have a much more difficult task to actually frame it in an evaluation phase.''} (H-1).

Some practitioners related the difficulty in UX evaluation to the fact that users' expectations and their perception of a product change over time and are affected by various factors; e.g., introduction of new technologies or appearance of a competitor's product.
In this regard, one of the interviewees highlighted:
\textit{``It's like when you try on new clothes. The shirt you were wearing going into the dressing room and looked fine, looks shabby when you've tried out the new shirt.''} (H-1).
The interviewee used this analogy to explain the subjective and dynamic nature of expectations, and that for each individual a new experience can affect the user's perception of other products.

The interviewees stated that there are still a number of open problem concerning UX evaluation.
One problem in evaluation concerns the temporality of UX and difficulties in relating the result of laboratory evaluations to the real experience of users (e.g. H-1).
Another problem concerns the relation between first impression of users (initial UX)  and overall UX or UX after using the software for a while (e.g. A-2).

\subsection*{C\ref{ch:lackcomp}.~Lack of consensus on UX-related competences and responsibilities}
Our data shows that to facilitate developing software with better UX, organizations need to have access to a variety of competences including brand management, visual design, usability engineering, interaction design, and emotional and pleasurable design.
The interviewees highlighted that their organizations still lack practitioners with competences for eliciting, refining, and specifying non-task-related or emotional needs of users.
This can explain the identified challenges concerning requirements work (C\ref{ch:re}) and evaluation (C\ref{ch:uxeval}).

Also, it is still not clear how organizations should manage this set of often quite differing competences.
The interviewees showed two main perspectives concerning this.
One group believes such a set of competences is hard to find in individual practitioners and all of the team members (with different competences) should take a joint responsibility towards UX.
Regarding this, one of the practitioners said:
\textit{``I'm not sure if that should be a specific role. [\dots] so everybody should have a UX focus now. I'm not sure if we can have some sort of UX guy [who takes the final decisions].''} (A-4).
He further argued that  achieving a better UX requires a `UX-mindset' in projects that even the most technical roles in the projects (e.g., programmers) should have.
The other group believes there is a need for specific individual practitioners with these multidisciplinary competences (i.e., defining specific UX-related roles).
Although in this case, often such practitioners find it difficult to `be a little bit of everything', and `juggle' their various competences (e.g. H-1).
Regarding the importance of individuals with diverse set of competences, another practitioner stated:
\textit{``I, as an art director, have to have  somewhat deep knowledge about UX, and also IxD \dots you can't separate them.''} (A-4).

The participants also discussed the responsibilities UX practitioners may have in projects, and tasks they should perform.
Our data shows that there is a direct relation between the extent of access to practitioners with various competences and responsibilities they take in projects.
Depending on the organizations studied, the UX practitioners had various responsibilities including performing user research, concept development, designing GUIs, testing usability of the software, requirements gathering (including both business and end user needs), creating design principles and guidelines, and performing market research.
As some UX practitioners expressed, they have varying responsibilities in and contributions to different projects; this depends on factors such as management support, available resources, and timing.
We also observed that the more technical practitioners are less informed about the responsibilities related to UX.

In general, our data shows that the responsibilities of UX practitioners are very diverse, and depend on various factors such as resources available, or the maturity of the company in relation to UX practices.
For instance, in one organization (C), the UX team is responsible for handling requirements and feeding them to the development teams.
One UX practitioners described her responsibility as:
\textit{``that can loosely be described as discovery, research, overall strategy, and then high-level design.''} (G-1).
However, in another organization (H), the UX group is part of  R\&D where the group mainly focuses on future products, and long term vision of the company.
As a parallel responsibility, the group gives feedback and support regarding UX to the development teams involved with current products.
In some organizations, since the number of practitioners with UX knowledge is low, none of these practitioners are part of any particular project teams, and are instead shared resources among them (e.g. H-1).
UX practitioners are also often responsible for spreading knowledge and awareness about UX in the organization.
Regarding this, a practitioner expressed:
\textit{``I think on a very high level, our responsibilities are to inform, influence and inspire.''} (H-1).
He further stated:
\textit{``we contribute to the process by running workshops, by providing provocative questions, or providing examples, engaging in discussions in which people from other domains dig down really deep into their own layers of knowledge, and we can ask really simple questions to poke them with our perspective.''} (H-1).

Being able to gain an overall view of UX design was one of the concerns highlighted in the interviews.
According to practitioners, this is often a difficult yet important prerequisite for creating a coherent UX.
As the interviewees stated, to deal with the complexity of today's systems, the common approach in the software community is to break down the whole system into various sub-systems and work independently on them.
Such an approach can harm the UX of the software since often in these cases UX practitioners lose the overall view on the UX design, how these different sub-systems fit together as a whole, and how they individually and in combination contribute to the experience of the end users.
Regarding this, one interviewee emphasized:
\textit{``you have to tear it down, yeah! but what happens to the whole? who is going to define the whole?''} (E-1).
The interviewees also highlighted that in agile processes, the decision-making process is more spread out both over time as well as among the team members.
This further complicates the process of creating a unified and coherent UX design (e.g H-1).
Further, agile processes enforce a focus on a few piece(s) of the design at each iteration.
Regarding this, one interviewee said:
\textit{ ``you need to deliver wireframes for parts of the application [features] but you still do not know how it all will fit together at the end.''} (A-2).

\subsection*{C\ref{ch:late}.~Late focus on UX in projects}
This challenge directly relates to competences and responsibilities (C\ref{ch:lackcomp}) but concerns `when and to what extent' these competences are put into practice, or how UX practitioners are involved in projects.

The interviewees generally agreed that early and continuously involving UX practitioners in projects is essential for achieving a better UX (e.g. A-4) and stated at least three benefits of that.
First, UX practitioners get first-hand information about the customer and end users.
Regarding this one UX practitioner stated:
\textit{ ``The worst case is when someone has met a client and talked a lot about the software, then I meet this guy who has met the client \dots   then it's secondhand information and everything gets distorted.''} (A-4).
Secondly, it is less likely that UX practitioners' input to the project is ignored.
Thirdly, different stakeholders get to discuss the trade-offs concerning UX design, for instance business versus user goals, or feasibility of design concepts considering the technical constraints.

The practitioners however highlighted that involving UX practitioners in earlier phases is not a substitute  for involving  technical practitioners.
In particular, the interviewees highlighted that early negotiation of trade-offs is of high importance for the success of the product in general, and UX in particular.
For instance, if required, the design concept can then be updated based on the developers' feedback.

Still, practitioners from 6 out of 8 companies stated they often come into projects only in later stages, and that this negatively affects their work.
The interviewees highlighted that still there is a common misconception among various stakeholders concerning UX- that it can be improved with just minor GUI changes in later stages of development.
While it is often impossible or difficult to make effective changes to the UX design in later stages.
One reason is that there is already a developed version of the software, or that and it is hard for developers to `kill their darlings' (H-1).
Also, the whole design concept might have been unsuitable, but a radical change at later stages may not be feasible due to time or effort constraints (e.g. H-2).
One of the practitioners highlighted that this is similar to how previously usability was treated in many organizations, as a `detection' and `fixing' step late in software development processes, and after the implementation was already done (e.g. H-1).

The group of UX practitioners highlighted that agile processes focus less on strategic decisions such as the overall UX of the software.
This means that often these strategic decisions, including decisions concerning UX, are either `skipped' or postponed.
Some practitioners related it to that agile methodologies tend to prioritize immediate and current problem solving:
\textit{``agile is a lot about problem solving and that's what sort of gets priority.''} (H-1).

According to the interviewees, even in cases when UX practitioners are involved in early phases, they may lose their connection to the project in later phases.
For instance some proposed design ideas maybe be changed by non-UX practitioners in later phases without consulting the UX practitioners (e.g. G-1).
A UX practitioner told us how the status of UX design is continuously checked during a project:
\textit{``we try to have always at least one person who was part of the original dialog present during the weekly checkups, and basically just going by the desks and checking, informally.''} (A-1).

\subsection*{C\ref{ch:communication}.~Communication and collaboration gap between UX and non-UX practitioners\footnote{Here, we use the terms `UX practitioner' and `non-UX practitioners' to respectively refer to  practitioners who have UX-related roles and responsibilities in the organizations and those who do not.}}
According to the interviewees, for various reasons UX practitioners face `power struggles' in relation to non-UX practitioners.
These include wide range of UX responsibilities (see C\ref{ch:lackcomp}), and that UX
should be taken into account from start to the end of a project (see C\ref{ch:late}).

As mentioned above, respondents argued that UX practices should start early in projects and that they impact all development phases, and different decisions in the projects.
This can be difficult for other project members to accept:
\textit{``sometime it can be perceived as we're trying to take control of the situation.''} (H-1).
According to the practitioners, this often means that working with UX is more difficult than usability:
\textit{``I think the reason why it was easier to work with usability to some extent was that you didn't take up any space. It was like being a woman in the early 20th century. You were there, but you didn't vote, you didn't do anything.''} (H-1)
Regarding the power struggle another practitioner stated:
\textit{``There are a lot of strong stakeholders that are really interested in doing those kind of things, programmers for instance, who like to be in control.''} (A-2).
He related this challenge to different motivations of these practitioners (e.g a developer wanting to develop more efficient code vs. a designer wanting to create a better design).
Similarly, another interviewee stated:
\textit{``In most companies if you have a go-ahead on a project or a research activity for instance, everyone wants to start with their own domain, as soon as possible, from day one. Then, of course, you have the problem of ownership of the direction, where to go and what to do and why. That's something that we struggle with quite a bit.''} (E-1).

In practitioners' views, one way to overcome this struggle is by informing other stakeholders about UX-related responsibilities (the `what', `why' and `how' of UX activities), especially in relation to the overall goals of the organization.
This shows that there is a clear relation between this challenge and C\ref{ch:kn} and C\ref{ch:value} where increased knowledge and awareness about UX and its value can contribute to addressing the power struggle between practitioners, at least to some extent.

According to practitioners, for a better UX practice,  there is a need for regular communication and collaboration among UX and non UX practitioners.
These two groups of practitioners often have different responsibilities, education, motivation, and constraints in their work.
Regarding the importance of communicating with developers, one UX designer stated:
\textit{``honestly the further we are away from the people that actually build the stuff, we run the real risk of becoming hand waving idiots.''} (G-1).
In his view, being disconnected from more technical roles, e.g., developers, runs the risk of not being aware of technological constraints when choosing and developing a design concept.
Another interviewee said:
\textit{``we really try to make this taking and giving. [\dots] we have constant communication, and I will say that  we get always input from developers that we need to consider.''} (A-2).
Similarly a developer said:
\textit{``I think we need to be working tighter with the design department to help them know what can be done.''} (D-1).

Regarding the importance of communication one practitioner highlighted:
\textit{`` The thing is that as human beings and organizations, the only place where just throwing something over the fence works really well is the military because you `have to' do it. Other than that, you have to build relationships. It is a social thing. If you ignore that aspect of the construction of anything, you're going to run into problems.''} (G-1).
She further emphasized:
\textit{``So be a responsible human being! Talk to the people that are eventually going to carry your work forward 'cause otherwise you're not going to be successful.''} (G-1).
Similarly another practitioner stated:
\textit{``I realized quite early in my career that I have to communicate with these guys who program or develop something, and I have to understand what they're saying.''} (A-4).

The practitioners however emphasized that overcoming this communication gap should be a two-way effort:
\textit{``I'm not saying that we should be the only ones with this kind of multidisciplinary approach. I think the other ones should also have that, [but] that's a big challenge, I would say.''} (H-1).
Similarly, another UX practitioner said:
\textit{``So I have to have some sort of knowledge about the technology because I have to know what my limitations are \dots. I have to have some sort of technical know-how so I can communicate with developers.
	I expect the same from them. So they realize that the aesthetic choice has to be made and it can take time.''} (A-4).

According to practitioners, in order to facilitate a better communication, UX practitioners need to acquire basic knowledge about various technical topics., e.g., programming, testing, architecture etc.
As emphasized by one respondent:
\textit{``You have to be like knowledgeable in many areas. \dots you have to be very holistic in the way you think about things, cause you have to speak with programmers in the language of programmers, to some extent, \dots. You also have to understand business.''} (H-1).
Similarly, another practitioner stated:
\textit{``you have to speak in engineer language.  That's a real challenge for UX, because you always have to translate it to terms that makes sense to an engineer or economist.''} (A-1).

Respondents, however, emphasized that communication between UX and non-UX practitioners can be challenging for various reasons including lack of trust in UX practitioners.
Regarding this, one interviewee emphasized:
\textit{``we have had problems with some of the developers sometimes.	It has been a bit of conflicts.
	I think we have some work to do to really to get a `we-feeling' that we,  together, are developing an awesome product.''} (C-1).
One practitioner related `lack of trust' to how the field of UX is a relatively new field and less established compared to more technical fields, e.g., SE.
This also means that UX practitioners can be often younger than non-UX practitioners.
\textit{``most UX practitioners are quite young still, I do not think they have been that long in the market for development of their competencies yet.''} (A-3).
One UX practitioner emphasized that they can gain more trust over time and as they accomplish more in their work:
\textit{ ``[over time] we are adding on the pile of what we would call successful things we have done, and of course that gives us a bit more `trust' I could say.''} (H-2).

\subsection*{C\ref{ch:customer}.~Customers resistance to the cost of UX practices}
One of the challenges in UX work that repeatedly came up in the interviews was that customers often resist taking on the costs of performing UX work (e.g. A-4).
Often customers believe UX is an add-on rather than a core concept that will help them improve their businesses.
This shows a clear relation between this challenge and C\ref{ch:kn} (concerning knowledge) and C\ref{ch:value} (concerning how UX is valued).

Regarding difficulties in convincing customers, a practitioner said:
\textit{``It can be hard. \dots If you have to have three weeks extra to make the graphics work in a certain way, they might think it's unnecessary because [software] will do fine. And maybe it's going to work fine, but if they made these extra efforts or put in this extra amount of money, they would have actually gone much, much further maybe.''} (A-4).
Similarly, a number of practitioners from a technology-focused organization emphasized that their customers are often too technology-oriented to care about UX (e.g. B-6).
On the other hand, it was not surprising to see that practitioners from the two UX consultancy companies found their customers more positive and open towards UX:
\textit{``Most people who invest in IT they want to succeed, and those who are hiring us they know that you have to know the users.''} (E-1).

As a way to convince customers to agree to the cost of UX practices, some practitioners use examples of successful products in the market (e.g., Apple) that are known for their good UX (e.g. A-4).
As another solution, one case company uses fixed prices for their projects so that practitioners can freely spend part of this money on UX practices (e.g. F-1).
Similarly, some practitioners emphasized that they talk about UX `indirectly'.
They connect UX to business goals, and argue for UX from the point of view of the business success, and not the end users.
An interviewee motivated this by saying:
\textit{``if you start babbling about usability and strange kind of things, they will say `oh! I don't want to pay for this'.''} (E-1).

According to some practitioners, UX practices should be sold to the customer as part of the contract to assure covering the associated costs.
This requires showing how such practices will add value to the software and have a return on investment (ROI) for the customer.
Nevertheless, often presenting a ROI is difficult.
This is evidenced by a UX practitioner who said:
\textit{``I don't think you can put ROI into a proposal necessarily. I think that's irresponsible, frankly. Because 95\% of the time, we don't understand the true and false nature of an issue when we're writing a proposal. It's only after working with a client for a little bit of time that we begin to see the nuance there. That usually undercuts any kind of understanding that's used to generate proposed improvements in ROI, for example.''} (G-1).

\subsection*{C\ref{ch:user}.~Low user involvement}
The interviewees highlighted that limited involvement of the end users in projects is another challenges to better UX work.
In their view, while customers should be involved to assure alignment of the projects to  business perspectives, the end users should be involved to assure alignment to UX-related needs (e.g. B-3).
This challenge can negatively influence requirements work (C\ref{ch:re})  and evaluation of UX (C\ref{ch:uxeval}).

We observed some inconsistencies in practitioners' views concerning user involvement.
While they all agreed on the importance of involving the end users, the view of more technical people regarding this matter was divided.
In favor of involving users an interviewee stated:
\textit{``maybe you want to have the end user involved also with the developers so that developers understand what they are doing, instead of just following the specifications. I think that would be very very valuable.''} (A-3).
On the other hand, a project manager stated that involving users in requirements discussions increases project costs.
Therefore, it is better to negotiate requirements and sign a contract without involving the users.
Regarding this an interviewee stated:
\textit{``[they] think they can say anything during [requirement] workshop and then get it. It is not the case. It is impossible for us to have this sort of infinity. So it leads to lots of long long long discussions afterwards.''} (B-2).
As another example:
\textit{``it usually leads to features that you take on more than what you agreed from the beginning. So it's possible that the customer gets a better system but they still don't pay you more money for this.''} (B-1).

One developer stated that they often have less access to end users, which can be problematic for their work mainly because this means developers do not often understand the rationale for requirements:
\textit{``developers they do not get that interaction [with the end users] cause they get their specification from the marketing people, and they get their specification from the interaction designers, but you don't get the motivation behind the requirements, because that gets lost during the way. So, the market people say that we must have this requirement, and interaction designer says we must do it this way, so you have the `what' and the `how' but you never get the `why'.''} (A-3).
Some practitioners emphasized relying too much on the end users' opinions might lead to less creativity in the design work:
\textit{``we have a quote sitting on the wall here, it's from Henry Ford [that says] `If I had asked people what they wanted, they would have said a faster horse'.''} (H-2).

	\subsection{Challenges Related to Each Other and Literature} 
	\label{sec:summary}

	Here we relate the challenges to existing literature and clarify their connections. 
	Even though, in our interviews, we did not focus on the connection between challenges, some were clearly stated. 
	However, future work is needed to detail and validate these connections.

	A limited or lack of knowledge of UX (C\ref{ch:kn}) can explain why it is not valued by individual practitioners or in organizations (C\ref{ch:value}).
	For example, some practitioners with more technical roles viewed UX simply as concerned with superficial changes to the GUI, not as a core concept.
	On the other hand, practitioners with a deeper knowledge of UX and how it differs from usability also thought it was more important and could provide a competitive edge. 
	This type of connection has been seen also in studies of other software quality characteristics. 
	For example, Berntsson Svensson et al.~\cite{BerntssonSvensson2012a} found if practitioners lack understanding and knowledge about software quality characteristics, they tend to undervalue, and ignore them during development.
	
	Similarly, a limited knowledge of UX (C\ref{ch:kn}) is likely also connected with the low impact of UX models in industry (C\ref{ch:models}).
	Several practitioners emphasized they do not work with UX by the book, or in an academic way.
	They seem to mostly gain their knowledge about UX from individual experiences rather than from existing UX theories and research. 
	This lack of a common reference and knowledge base can lead to that the concepts of UX and UX practices mean different things to different people, i.e., a language gap.
	This gap may contribute to communication problems within software development organizations.
	Communication problems, as empirical studies show, may lead to failure to meet the customers' expectations and quality matters~\cite{Bjarnason2011}.
	
	UX models can help overcoming the language gap by providing a shared understanding, and definition of UX.
	However, UX models for industry use are still rare as our interviewees stated (C\ref{ch:models}) and previous research emphasizes as well~\cite{Folstad2010a}.
	We find at least four explanations for low industrial impact of the current UX models: 
	(i) they are not standardized yet or included in main software development and engineering textbooks,
	(ii) they do not represent the relation between UX and other more established software quality models (e.g., ISO/IEC~25010~\cite{ISO250102011}),
	(iii) they are developed outside the field of software development and therefore often use terminologies less known to researchers and practitioners in this field,
	(iv) they are not supported by guidelines on how to apply them in practice, especially in combination with existing software quality characteristics models.
	
	Another connection between challenges concerns software companies' lopsided focus on functionalities and objectively measurable quality characteristics (C\ref{ch:techfocus}) and problems with requirements work (C\ref{ch:re}) and testing (C\ref{ch:uxeval}).
	Our findings show that practitioners focus more on functionalities or other software quality characteristics, and often perform UX-related requirements work and testing either informally, or abandon them in time and budget pressure.
	In addition, even if companies intend to focus more on UX (C\ref{ch:techfocus}), they do not always have the capability to turn this intention into action.
	For instance, they still lack required competences (Ch\ref{ch:lackcomp}) or tools and methods  (C\ref{ch:models}).
	This type of connection has been discussed also in previous studies.
	For example, Ardito et al.~\cite{Ardito2014b} argue that if practitioners fail to include UX and usability in requirements documents, these requirements might be ignored in testing.
	Chung et al.~\cite{Chung2009a} stress that researchers and practitioners traditionally have focused more on functionalities therefore developed requirements models or specification languages that do not sufficiently support quality characteristics.

	Challenges in requirements work (C\ref{ch:re}) and testing (C\ref{ch:uxeval}) are also connected with challenges in involving UX practitioners in projects.
	It is more difficult to identify, refine, document, and test UX-related requirements and companies still lack required competences (Ch\ref{ch:lackcomp}).
	Even if UX practitioners with the right competence are present in a company, they often face power struggle, and lack of trust therefore fail to effectively communicate and collaborate with non-UX practitioners (Ch\ref{ch:communication}).
	Isomursu et al.~\cite{Isomursu2012a} also show that there is a lack of effective communication and collaboration between UX and non-UX practitioners in agile projects.		
	They further emphasize that this lack leads to difficulties in integrating UX practices with other development activities.

	Difficulties in requirements work (C\ref{ch:re}) and testing (C\ref{ch:uxeval}) were frequently expressed in the interviews.
	Similarly, literature on software quality characteristics reports that practitioners find it more difficult to perform requirements and testing activities for these characteristics than for functionalities~\cite{Chung2009a, Paech2004a,BerntssonSvensson2012a}.
	Researchers have long emphasized addressing such difficulties, and argued that failing to do so can lead to undervaluing, or abandoning these characteristics in software projects, and consequently failing to deliver quality software~\cite{Chung2000a,Paech2004a}.
	Considering the similarities between UX and other software quality characteristic, it is reasonable to assume that the same consequences can be observed if challenges in UX-related requirements work and testing are not addressed.	
	
 	\section{Discussion and Analysis}
	Based on \intervieweesnum interviews in industry, we identified \challengesnum challenges that practitioners face when trying to integrate UX in their software companies.
	Some of these challenges have been mentioned in previous studies on UX.
	However, our study is broader: we consider more contexts and find a larger set of challenges.
	Similar challenges to those expressed by our interviewees are in fact reported in literature on usability~\cite{Rosenbaum2000a,Boivie2003b,Jerome2005a,Heiskari2009a,Gulliksen2004a} and/or other software quality characteristics~\cite{Chung2000a,Chung2009a,Paech2004a,BerntssonSvensson2012a,Borg2003a}.
	Nevertheless, the participants of our study either view many of the challenges unique to UX, or more severe than for usability or other quality characteristics.
	For example, practitioners find it challenging to document measurable usability requirement~\cite{Gulliksen2004a,Cajander2013a}, the same goes for other quality requirements~\cite{BerntssonSvensson2012a,Borg2003a}.
	Nevertheless, the interviewees emphasized they experience extra difficulties in documenting UX-related requirements.
	
	Similar to our interviewees, current UX studies seem to present challenges as unique or more difficult for UX as well.
	Often, researchers that study the practice of UX, and explicitly differentiate it from usability~\cite{Lallemand2014a,Larusdottir2012a,Law2015a,Lanzilotti2015a} do not compare and contrast UX challenges to similar ones reported for other software quality characteristics or even usability.
	Although one reason might be the study scope, this can also indicate that these researchers, if not deliberately, accidentally report these challenges as unique to or extra difficult for UX.
	However, there are exceptions:
	Vermeeren et al.~\cite{Vermeeren2010a} compare challenges in evaluating usability and UX, and discuss extra difficulties practitioners face in UX work, e.g how to evaluate UX in earlier phases when no functional software exists to interact with.
	Similarly, Isomursu et al.~\cite{Isomursu2012a} discuss that lack of tools and methods to objectively measure UX adds extra difficulties to the practice of UX compared to usability.
	Still, these exceptions provide few details or explanation for these extra difficulties.
	Current studies often associate the challenges to the gaps  (e.g cultural, educational and terminology gaps) between the two fields of HCI and SE because the concepts of UX and usability originated in the field of HCI and outside SE~\cite{Seffah2005a-ch1}.
	Nonetheless, similar challenges to the ones reported in those studies are even experienced inside the field of SE and for other quality characteristics local to SE (e.g reliability, and security).
	One example is how too much focus on functionalities can lead to challenges in usability work.
	While often usability studies assign this to different cultures of developers versus usability professionals~\cite{Seffah2005a-ch1}, similar pattern is seen for other quality characteristics as well.
	For instance, Berntsson Svensson et al.~\cite{BerntssonSvensson2012a}  empirically show that too much focus on functionalities can lead to abandonment of quality characteristic in projects.
	Therefore, we conclude that the gaps between the fields can partially explain the presence of challenges but not the overlaps.
			
	We acknowledge the observed overlaps but still trust our data that shows practitioners often face extra difficulties in working with UX.
	Therefore, we propose that despite similarities at a superficial level, we can differentiate the challenges at a deeper level, through two unique aspects of UX: its \textit{subjectivity} and \textit{emergent nature}.
	Experience is a holistic, temporal phenomenon emerging from its underlying, intertwined elements (i.e., subjective human perception, action, motivation, emotion, and cognition)~\cite{Hassenzahl2010a}.
	For instance, one user may perceive particular features of the software as simple, novel, and admirable while another one may perceive the same features as complicated and old.
	In addition, a user's perception, or emotion is likely to change over time and as the user interacts more with a piece of software.
	For example, over time, the user may find the same novel features as old, or a complex feature as simple.
	Today, we do not still have much understanding of how  users' perceptions change over time~\cite{Hassenzahl2003a}.
	In contrast to UX, other quality characteristics are less dependent on user perception or time. 
	For instance, performance or security measurements will have the same results even when repeated over time, providing that the software, and the test context (e.g., CPU load) have not changed.
	A user, however, might for instance \textit{perceive} the same software as secure in the first test but insecure in the second, because she has heard a similar software has been recently hacked.
	Also, UX is formed through \textit{a complex interrelation between its underlying elements}.
	This resembles the cross-cutting nature of other quality characteristics.
	In both cases, more than one functionality is affected by the related requirements.
	For instance, a designer can select a group of specific functionalities in order to increase the likelihood of evoking a particular emotion in the end users.
	However, in case of UX, this interrelation is more complex compared to other quality characteristics.
	Hassenzahl~\cite{Hassenzahl2006b} emphasizes that UX is an emergent phenomenon that is not totally predictable, or reducible to its underlying elements.
	Therefore, there is no guarantee that a designer can assure creating a certain experience through a set of design choices~\cite{Hassenzahl2010a}.
	
	In relation to being subjective and emergent, we see at least \issuesnum issues that are essential to the very nature of UX (i.e., essential issues)\footnote{The terms essential and accidental in analysing underlying causes were originally used by Aristotle, and later adopted in the context of software development by Brooks~\cite{Brooks1987a} in his classification of complexities in software engineering.}. 
	Essential issues can only be addressed partially and will therefore likely continue to impact the practice of UX by adding various extra difficulties to the work of practitioners.
	These issues can explain why practitioners perceive UX challenges to be compounded.
	In total we identified \difficultiesnum  such difficulties in relation to the essential issues, summarized in Table~\ref{tbl:issues} and discussed below.
	Clearly, our study did not focus on identifying these issues or difficulties, and we require further research to provide a more comprehensive understanding of them, and their interconnections.

	One issue is that practitioners often need to involve \textit{statistically significant number of heterogeneous users} to guarantee reliable results when measuring UX~\cite{Law2014b}.
	UX relies on perception and emotion of users, and currently, the most efficient and feasible approach to  measure them  is to directly gather users' opinions, and let the users express themselves~\cite{Law2014b}.
	This is often performed through questionnaires or scales (e.g., AttrakDiff, Self-Assessment Manikin, Affect Gird~\cite{Zimmermann2008a}).
	However, to gain reliable results, practitioners need to gather responses from statistically significant number of heterogeneous users~\cite{Law2014b}.
	Nevertheless, practitioners have long struggled with limited access to users~\cite{Rosenbaum2000a} as our findings confirm as well.
	This limited access can negatively impact UX measurement for example practitioners may gather their peers' opinions rather than users'~\cite{Larusdottir2012a}.
	Clearly, in such cases, the measurement results do not necessarily reflect users' perception.
	In contrast to UX, practitioners can test other software quality characteristics even without involving users (i.e., automatically).

	Another issue is that for measuring UX, practitioners often need to use \textit{sophisticated prototypes} to be able to gather authentic experiences and perceptions of users~\cite{Law2014b,Vermeeren2010a,Zimmermann2008a}.
	For example, practitioners in Law et al.'s study~\cite{Law2014b} emphasized they need to use variety of media (e.g., video, TV, social media) to develop the required prototypes for measuring UX and that they often even need more than one such prototypes to gather enough input for design.
	However, often practitioners do not have access to such prototypes especially in earlier phases of software development; which can lead to ignoring UX measurement in these phases~\cite{Vermeeren2010a}.
	One solution is to develop methods that do not require such prototypes but can for instance rely on practitioners' imagination of how users may perceive the design~\cite{Vermeeren2010a}.
	This is however still an open research topic.
	In contrast to UX, practitioners can test other software quality characteristics even without necessarily using a sophisticated prototype of the software.
	For example, usability can be tested on simple paper prototypes, or performance problems can be avoided via early modeling of the architecture~\cite{Balsamo2004a}.
	
	Another issue is that UX measures are essentially \textit{prone to fading and fabrication}~\cite{Law2014b} mainly because users' memory of their experience can easily change.
	This also means that UX measures are \textit{highly sensitive to timing and nature of tasks}~\cite{Law2014b}.
	Based on length and complexity of a task, practitioners can decide to measure emotions during or after it is completed, because a user may feel differently in different stages of performing  the task.
	In addition, what a user remembers about her emotions may not reflect the reality.
	As one of our interviewees stated `they will try to explain it but it perhaps is not the real emotion that we catch at the end.'
	Therefore, practitioners need to decide when and how to measure the overall UX, or its underlying elements (e.g., emotions).
	This is difficult mainly because the community still lacks enough understanding of the relation between UX, time and memory, and suitable UX metrics and measures that can sufficiently support this relation~\cite{Law2014b}.
	For example, empirical data shows that practitioners still  have no means to measure the exact emotion of users at each moment~\cite{Law2014b}.
	This agrees with our data, our interviewees also stated that `it is a feeling thing'  or that although `it is difficult to measure emotions'.
		
	Another issue is that UX is not totally reducible to its complexly intertwined underlying elements~\cite{Hassenzahl2010a}.
	This is even acknowledged by those researchers that model UX through a composition of elements (e.g., task-related and non-task-related~\cite{Hassenzahl2003a}).
	These researchers emphasize  that however practitioners may `manipulate' UX of a product through these elements, they cannot `guarantee' a certain overall UX by doing so~\cite{Wright2010a,Hassenzahl2010a}.
	Consequently, practitioners and researchers have different views on how UX should be measured and whether they can predict the overall UX of a piece of software by merely measuring these elements~\cite{Law2014b}.
	For example, Law~\cite{Law2010a} argues that often there are two groups of practitioners and researchers with different attitudes towards UX measurement: those who are strongly convinced that it is  ``necessary, plausible and feasible'' to measure UX through its finest underlying elements, and  those who are doubtful about the ``necessity and utility''  of measuring the elements.
	This is compounded by the fact that the community still lacks a clear understanding of how UX underlying elements interact and influence each other~\cite{Hassenzahl2006b}.
	For example, empirical data shows that users perceive a piece of software to be more usable when it is more beautiful~\cite{Hassenzahl2010a}.
	In addition, practitioners do not still have enough guidelines on how to choose suitable UX measures and metrics for UX underlying elements, and interpret their findings to improve the overall UX~\cite{Law2014b}.
	Further, it is not still clear how practitioners should pick underlying elements to increase the likelihood of delivering a certain overall UX, and there is a general lack of guidelines, tools and methods for that purpose.
	For example, our interviewees stated that it is hard to `merge these requirements into design'.
	They also stressed that UX and non-UX practitioners still disagree on whether they should identify and refine UX abstract requirements in parallel, before or after other functional and quality requirements.
	
	 Researchers highlight that  since quality characteristics are hard to measure, practitioners tend to judge them based on their personal opinion~\cite{Paech2004a,Chung2000a}; thus they are generally known to be subjective.
	This is similar to our observations about UX that practitioners also face difficulties in measuring UX.
	Therefore, it is expected to see that they use their own subjective perception rather than users' perception to judge the level of UX of the software.
	Naturally, practitioners can have different views on whether the requirements are satisficed or not.
	That can therefore lead to disagreements among the stakeholders.
	Thus, to overcome subjectivity and such disagreements, practitioners are recommended to document requirements related to these characteristics in a measurable manner~\cite{Chung2000a}.
	For that purpose, practitioners need to have access to suitable, and agreed upon metrics and measures.
	However,  this is more difficult in case of UX mainly because as we have seen, we still lack enough knowledge and consensus regarding these measures and metrics.
	While UX metrics and measures are not agreed upon or standardized yet, for other quality characteristics practitioners have access to standards, e.g., ISO/IEC~9126-4~\cite{ISO9126-4-2004}.
	Thus, compared to other quality characteristics, subjectivity of UX is harder to tackle.
	
	In summary, some of the challenges we identified overlap with those reported in existing literature about usability or software quality characteristics.
	However, through \issuesnum essential issues that relate to  the \textit{subjective} and \textit{emergent} nature of UX, we explained the extra difficulties practitioners face in UX work compared to other software quality characteristics.
	These issue also differentiate subjectivity and emergent nature of UX from  \textit{subjectivity} of software quality characteristics and their often \textit{cross-cutting} nature.
	Although these issues are essential to the very nature of UX, and cannot be totally overcome, researchers and practitioners should take them into consideration in particular when developing tools, methods and guidelines to overcome the challenges.
	Our findings can be useful for researchers in identifying new and industrially relevant research areas, and for practitioners who want to learn from empirically investigated challenges in UX work, and base their improvement efforts on such knowledge.
	By highlighting the identified overlaps between UX challenges and previously known challenges in literature on software quality characteristics we can help finding research areas useful not only in improving the practice of UX but also software quality in general.
	They can also make it easier for practitioners to spot, better understand, as well as find mitigation strategies for UX, through learning from past experiences and developments in the area of software quality.
	
	\begin{table}
		\footnotesize
	\begin{tabularx}{\textwidth}{|X|X|X|}
			\hline
\textbf{UX is subjective} \& \textbf{emergent}	& \textbf{Essential Issues} & \textbf{Extra difficulties }
\\ \hline
\begin{itemize}[topsep=0pt,leftmargin=*]
	\item UX relies on human subjective perception
	\item  UX is temporal
	\item  UX emerges from complexly intertwined underlying elements
\end{itemize}
	&
	\begin{itemize}[topsep=0pt,leftmargin=*]
		\item UX measures are sensitive to timing and nature of tasks
		\item  UX measures are prone to fabrication and fading (i.e., impact of memory)
		\item  UX measures require statistically significant num of users
		\item  UX measures require sophisticated prototypes 
		\item  overall UX is not totally predictable or reducible to its underlying elements
	\end{itemize}
  & 
  \begin{itemize}[topsep=0pt,leftmargin=*]
  	\item a deeper understanding of the relationship between UX and time is missing
  	\item  a deeper understanding of the relationship between UX and memory is missing
  	\item  practitioners do not always have access to enough users
  	\item  practitioners do not always have access to sophisticated prototypes in earlier phases
  	\item  practitioners do not often know how to measure the whole UX in relation to its underlying elements 
  	\item  practitioners do not always have access to standardized  and agreed upon set of UX measures and metrics
  	\item abstract quality characteristics (e.g., non-task-related or emotional ones) are difficult to refine and translate into more concrete ones
  \end{itemize}
\\ \hline
\end{tabularx}
\caption{We identified \issuesnum issues essential to the very nature of UX (in particular its subjective and emergent nature).
	These issues can explain the extra difficulties that practitioners face in their work with UX compared to other quality characteristics}
\label{tbl:issues}
\end{table}

	\section{Conclusion}
Our work answers calls for more empirically based studies on the practice of UX in particular, and software quality characteristics in general.
We provide  an increased understanding of UX challenges  in the software industry, that can help the community to identify ways to systematically improve the current practice of UX.
We show that software development practitioners have a large number of challenges in integrating UX practices into their development processes and organizations; in total we identified \challengesnum challenges as mentioned in \intervieweesnum interviews in eight software organizations.

Our findings and analyses enhance previous empirical studies on challenges by both corroborating previously found empirical evidence and by providing a deeper explanation and understanding of the nature of these challenges.
We also realized that subjectivity and emergent nature of UX can explain the extra difficulties practitioners face in the practice of UX compared to other software quality characteristics and in particular usability.
Despite the similarities these aspects have to subjectivity of other quality characteristics or their cross-cutting nature, we realized that at least \issuesnum essential issues differentiate UX from other characteristics: UX measures are prone to fading and fabrication and sensitive to timing and nature of tasks, they require statistically significant number of users and sophisticated prototypes, UX is not totally reducible to its underlying elements.

In conclusion, to make future progress in integrating UX practices into software development processes, the community needs to take  these essential issues into account when developing guidelines, tools and methods  to address related challenges.
More importantly, our findings can shed light on how practitioners should integrate less mature and new knowledge areas such as software quality characteristics and UX  into software development processes.

\section*{Acknowledgments}
The authors would like to thank all the organizations and in particular practitioners who participated in this study.

\section*{References}

\bibliographystyle{IEEEtran}

\bibliography{\libpath}

\section*{Appendix A: Interview guide}

\subsection*{General Questions}
\begin{itemize}
\item  What is your education and work background?
\item  What is your role in this company?
\item How many years have you had this role?
\item Do you know any of these terms (see Appendix C)? if yes, how do you apply them in your work?
\item How do you define UX?
\item How do you define Usability?
\end{itemize}

\subsection*{Questions Related to Requirements}

\begin{itemize}

\item How is the overall requirements process in your company?

\item How do you approach functional requirements?

\item How do you approach non-functional/quality requirements?

\item  How do you approach requirements related to UX?

\item What challenges do you face in your work regarding requirements related to UX?

\end{itemize}

\subsection*{Questions Related to Design}

\begin{itemize}
\item How is the overall design process in your company?

\item How is `design' related to `requirements' in your work?

\item What challenges do you face in your work regarding design, in particular in relation to UX?

\end{itemize}

\subsection*{Questions Related to Evaluation or Testing}

\begin{itemize}
\item How is the overall evaluation/testing process in your company?

\item How do you test functional requirements?

\item How do you test non-functional requirements?

\item  How do you test UX?

\item What challenges do you face in testing UX, or requirements related to UX?

\end{itemize}

\section*{Appendix B: Coding guide}
\begin{itemize}

\item Every segment can have any number of applicable codes

\item The codes should be selected from the list below.
If  a new concept appears in data the possibility of adding a new code should be discussed among the authors.

\item Any uncertainty in coding a segment should be discussed among the authors

 \subsection*{List of Codes}
\begin{enumerate}
\item{Challenges}
\item{Solutions}
\item{UX}
\item{Usability}
\item{UX vs Usability}
\item{Motives}
\item{Definition}
\item{Organization}
\item{Project}
\item{Software}
\item{Process}
\item{Individuals}
\item{Tools and methods}
\item{Roles}
\item{Responsibilities}
\item{Collaboration}
\item{Communication}
\item{Requirements}
\end{enumerate}
\end{itemize}

\section*{Appendix C: Terminology Table}
The interviewees were asked to specify each and every term they know and whether they apply it in their work.
They were also asked to add any relevant term that is missing from the list.

\begin{itemize}
\item Usability
\item User Experience
\item Quality in Use (QiU)
\item Emotional design
\item Pleasurable design
\item Aesthetics of design
\item Affective computing
\item Affective design
\item Usability requirements
\item UX requirements
\item Affective requirements
\item Emotional requirements
\item User values
\item User emotions
\item User motivations
\item ISO/IEC~9126
\item ISO/IEC~25010
\item Hedonic and pragmatic
\item Instrumental and non-instrumental
\end{itemize}

\end{document}